\documentclass[11pt,paper]{article}
\usepackage{jheppub}

\usepackage{float,slashed}
\usepackage[usenames,dvipsnames,table]{xcolor}
\usepackage{graphicx,amsmath,amssymb,multirow,array,bm,mathrsfs}
\usepackage{epsf,amsfonts,slashed}
\usepackage[numbers,sort&compress]{natbib}
\usepackage[export]{adjustbox}
\usepackage{tikz}
\usetikzlibrary{decorations.pathmorphing}
\usetikzlibrary{decorations.markings}
\usepackage{scalerel,stackengine} %Used to define a long hat
\usepackage{soul}
\usepackage[utf8]{inputenc}
\usepackage{hyperref}
\usepackage{tikz-feynman}

\stackMath

\title{
Polyakov's confinement mechanism 
for\\ 
generalized Maxwell theory
}

\preprint{PUPT- 2636}

\author{Matthew Heydeman,${}^{a,b}$ Christian B. Jepsen,${}^c$ Ziming Ji,${}^{d,e}$}
\author{Amos Yarom${}^f$}

\affiliation{$^a$School of Natural Sciences, Institute for Advanced Study, Princeton, NJ 08540, USA}
\affiliation{$^b$Joseph Henry Laboratories, Princeton University, Princeton, NJ 08544, USA}
\affiliation{$^c$Simons Center for Geometry and Physics, Stony Brook University, Stony Brook, NY, 11794}
\affiliation{$^d$SISSA, Trieste, 34136, Italy}
\affiliation{$^e$INFN, Sezione di Trieste, Via Valerio 2, 34127 Trieste, Italy}
\affiliation{$^f$Department of Physics, Technion, Haifa 32000, Israel}

\emailAdd{mheydeman@ias.edu}
\emailAdd{cjepsen@scgp.stonybrook.edu}
\emailAdd{zji@sissa.it}
\emailAdd{ayarom@physics.technion.ac.il}

\abstract{
We study fractional-derivative Maxwell theory, as appears in effective descriptions of, for example, large $N_f$ QED${}_3$, graphene, and some types of surface defects. We argue that when the theory is UV completed on a lattice, monopole condensation leads to a confining phase via the Polyakov confinement mechanism.%}
}
\begin{document}

\maketitle

\section{Introduction and summary}
In the study of infrared effective field theories, it is often useful to integrate out massive degrees of freedom. Contrariwise, it is not common to integrate out massless degrees of freedom, mainly because integrating them out generates non-local terms in the effective action. These obstacles notwithstanding, there are theories for which this practice has proved fruitful. One example is QED$_3$ with a large number $N_f$ of massless fields, which when integrated out produce a non-local photon propagator \cite{Appelquist:1981vg, Appelquist:1988sr} (see also \cite{Anselmi:2000fr, Borokhov:2002ib, Witten:2003ya,Giombi:2015haa,Chester:2016ref,Giombi:2016fct} for a handful of related modern discussions). Another example involves integrating out bulk fields in the semiclassical limit of the AdS/CFT correspondence \cite{Maldacena:1997re,Gubser:1998bc,Witten:1998qj}, which leads to boundary generalized free field theories. References \cite{Vasiliev:1990en,Klebanov:2002ja,Witten:2003ya, Giombi:2012ms,Giombi:2013yva} studied $N_f$ QED$_3$ and similar vector models in this holographic setup. Related examples include theories on manifolds with a boundary where bulk field are effectively integrated out, e.g., free bulk scalar fields which interact on the boundary \cite{Paulos:2015jfa,Herzog:2017xha,Giombi:2019enr}, quantum systems interacting with an environment \cite{PhysRevLett.46.211,Callan:1989mm,Callan:1994ub}, and the theory of four-dimensional photons that couple to electrons restricted to live on a three-dimensional surface---a system which admits a purely three-dimensional effective description involving a non-local Maxwell term \cite{Marino:1992xi,Teber:2012de,Teber:2014ita,Herzog:2017xha,Karch:2018uft,Dudal:2018pta,DiPietro:2019hqe,Herzog:2020bqw,Pedrelli:2020zlw, Heydeman:2020ijz, Herzog:2022jqv}, and which has been usefully applied as a model of graphene \cite{Semenoff:2011jf,Teber:2014ita}.

Motivated by the above considerations, consider the generalized free Maxwell theory in three dimensions,\footnote{We refer the reader to Ref. \cite{frasca2021confinement} for a recent study of the RG flow and confinement of non-local gauge theory in four dimensions, although there non-locality arises from an exponentiated, rather than a fractional, Laplacian.} 
whose action is given by 
\begin{align}
\begin{split}
\label{E:U1action0}
    S&=\frac{1}{4e^2}\int_{\mathbb{R}^3} d^3x\, F_{\mu\nu}D^{s-2}F^{\mu\nu} \\
    &=\frac{1}{4e^2}\int_{\mathbb{R}^3} \frac{d^3k}{(2\pi)^3}\, \widetilde{F}_{\mu\nu}(-k)|k|^{s-2}\widetilde{F}^{\mu\nu}(k)\,,
\end{split}
\end{align}
where tilde'd quantities are Fourier transformed quantities,
\begin{equation}
\label{E:Fourier}
	\widetilde{\phi}(k) = \int d^dx\, \phi(x)\, e^{ik x}\,,
\end{equation}
$D^{s-2}=(-\nabla^2)^{(s-2)/2}$ is the fractional derivative operator (which is defined through the second equality in \eqref{E:U1action0}), $s$ is a real number between zero and two, and $F_{\mu\nu}=\partial_\mu A_\nu-\partial_\nu A_\mu$ is the field strength associated to a $U(1)$ gauge field. We restrict ourselves to flat space throughout this paper but refer the reader to \cite{basteiro2022fractional} for recent work on fractional-derivative theories in curved space.

In what follows, we will focus primarily on the case $s=1$, where \eqref{E:U1action0} corresponds precisely to the non-local kinetic term appearing in large $N_f$ QED and the effective theory of graphene mentioned above. Thus, when $s=1$ we may think of 
\eqref{E:U1action0} as an effective action of an ultimately local theory. As such, we may take the gauge field to satisfy the Dirac quantization condition. If we set $s=2$, we recover standard three-dimensional Maxwell theory with a compact gauge group. 

When $s\neq 1$ and $0<s \leq 2$ the theory cannot be obtained by integrating out modes of a local theory in an integer number of dimensions. Nevertheless, theories with a non-local kinetic term identical, or similar to that in \eqref{E:U1action0} appear throughout the literature. For instance, the continuum limit of the long range Ising model possesses a non-local kinetic term, $D^s$, \cite{PhysRevLett.29.917,Sak,Honkonen:1988fq,Honkonen:1990mr,Koffel:2012cu,Behan:2017dwr,Behan:2017emf,Benedetti:2020rrq} and can be related to integrating out a free bulk scalar field for appropriate values of $s$ \cite{CStheorem,Paulos:2015jfa}. Similarly, certain aspects of turbulent flow have been tied to field theories with non-local kinetic terms in \cite{Oz:2017ihc,Levy:2018xpu,Levy:2019tjl,Kislev:2022emm}. More recently, theories of this type have been shown to posses a symmetry-broken phase, persistent at high temperatures \cite{Chai:2021djc,Chai:2021tpt}. The theory in \eqref{E:U1action0} for arbitrary $s$ has been related to models of superconductivity in \cite{LaNave:2019mwv} and has been studied extensively in \cite{Kleinert:2002qk,Herbut:2003bs}. More general properties of field theories with a non-local kinetic term have been considered in \cite{doAmaral:1992td,Marino:2014oba,Basa:2019ywr,calcagni2021quantum,heredia2022nonlocal,calcagni2022ultraviolet}.

Theories with non-local kinetic terms, of the type exemplified by \eqref{E:U1action0}, raise many questions regarding the viability of the associated effective theory. For instance, one may wonder whether it maintains unitarity or causality, or, what type of initial value problem classical solutions satisfy.  The theory in \eqref{E:U1action0} is free and unitary, but once interactions are introduced the resulting theory is expected to break unitarity \cite{Heydeman:2020ijz}. This general expectation relies on a violation of the optical theorem in an effective boundary theory generated by integrating out bulk modes.

The question of causality was addressed in Ref. \cite{doAmaral:1992td}, which studied the quantization and causality of generalized Maxwell theory and found that causality is respected in the sense that the classical Green's function vanishes outside the light cone. The approach to canonical quantization adopted in this reference consisted in introducing an auxiliary mass term $\chi$ and performing an expansion of a fractional derivative $D^\sigma$:
\begin{align}
(-\nabla^2)^{\frac{s-2}{2}}
\rightarrow 
(-\nabla^2+\chi)^{\frac{s-2}{2}}
=\chi^{\frac{s-2}{2}}
-\frac{s-2}{2}\chi^{\frac{s-2}{2}-1}\nabla^2
+\frac{(s-2)(s-4)}{8}\chi^{\frac{s-2}{2}-2}(\nabla^2)^2
+
...\,,
\end{align}
which results in infinitely many derivatives. Since each term is local, Hamilton's formalism applies, but there are no independent momenta. 

In general, whether an infinite-derivative theory is a well-behaved or not depends on whether the coefficients of the higher-derivative terms decay sufficiently fast with the number of derivatives. The best kind of scenario is illustrated by Ref. \cite{kleppe1992nonlocal}, which studied a modified Yang-Mills theory where the propagator was dressed with factors of $\exp(-\nabla^2/\Lambda)$ and showed that on-shell tree-amplitudes are unaffected by this regularization procedure. A related result was reported in Ref. \cite{li2011holography}, wherein it is showed that for theories with an exponentially decaying tower of derivatives, the zero-temperature entanglement entropy exhibits the usual area law scaling; in Ref. \cite{Basa:2019ywr} it was further demonstrated that such theories can be me mapped onto purely local theories. In contrast, however, for fractional-derivative theories like generalized free Maxwell theory, Refs \cite{li2011holography,Basa:2019ywr} explicitly find that this local behaviour is not present.

But while fractional-derivative theories cannot immediately be lumped together with local theories, there are also issues with formally treating them as infinite derivative theories, especially as far as the classical equations of motion are concerned. The problem with an infinite-derivative theory of a field $\phi(x)$ is that it has an infinite-derivative equation of motion. So for the initial conditions, one needs to stipulate not just $\phi(0)$ and $\phi'(0)$, but all the higher derivatives as well, $\phi''(0)$, $\phi'''(0)$, etc. But if all the derivatives of a function at a given point are specified, then, up to issues of non-analyticity, the entire function is determined.

Ultimately, fractional-derivative theories are not the same kind of beasts as infinite-derivative theories. We can see this by studying the equations of motion. In Fourier space the local ($s=2$) equations of motion resulting from \eqref{E:U1action0} read
\begin{align}
	0 = k^\mu \widetilde{F}_{\mu\nu}(k)\,,
\label{localFourier}
\end{align}
while the non-local equations of motion read
\begin{align}
	0 =\frac{k^\mu\widetilde{F}_{\mu\nu}(k)}{|k|^{2-s}}\,.
\label{nonlocalFourier}
\end{align}
It may be tempting to conclude that these two equations are equivalent, but that would be too hasty. For example, Laplace's equation $\nabla^2\phi =0$ in $\mathbb{R}^d$ Fourier transforms to $0=k^2\widetilde{\phi}$, but this does not imply that $0=\widetilde{\phi}$. There are solutions where $\widetilde{\phi}$ is a generalized function with support only at $k=0$. Similarly, there are solutions to \eqref{nonlocalFourier} that are not solutions to \eqref{localFourier}, namely solutions for which $k_\mu \widetilde{F}_{\mu\nu}(k)$ is a generalized function with support only at $k=\infty$. Once we discard such pathological solutions as not belonging to the space of functions over which the path integral for generalized free Maxwell theory should be performed, then the local and non-local equations of motion equivalent.

While its kinematics are non-local, the field strength $F$ is gauge invariant.
Due to the Bianchi identity, the current $J^\mu= \epsilon^{\mu \nu \rho} F_{\nu \rho}$ is automatically conserved, implying that the family of non-local free theories possess, in the language of \cite{Gaiotto:2014kfa}, a generalized global magnetic $0$-form symmetry. 
In \cite{Heydeman:2020ijz}, it was shown that for $s=1$, \eqref{E:U1action0} is conformally invariant and $J$ is a conserved current in this CFT. 
The existence of this topological current corresponds to the presence of magnetically charged operators, the monopole operators. In radial quantization on the cylinder $\mathbb{R}\times S^2$, a monopole operator corresponds to a state on $S^2$ with non-vanishing magnetic flux,
\begin{equation}
    \int_{S^2} F = 2 \pi q \, ,
    \label{E:monopoleflux}
\end{equation}
where $q\in \mathbb{Z}$. The importance of magnetic sources in three dimensions was first emphasized by \cite{Polyakov:1975rs}, and later work established these monopoles as local operators in conformal field theory (see \cite{Borokhov:2002ib,Pufu:2013vpa,Dyer:2013fja,Chester:2016wrc}.)
As it stands, \eqref{E:U1action0} is a free theory with no magnetically charged sources, so there is no dynamical mechanism to study the effects of monopoles. We will remedy this by embedding \eqref{E:U1action0} in a bigger UV theory which does have monopole solutions which contribute to the path integral as instantons.

Indeed, in his famous paper \cite{Polyakov:1975rs}, Polyakov argued the $s=2$ theory exhibits the phenomenon of confinement once monopole solutions are allowed to proliferate in the path integral. One way to achieve this is to show that 3$d$ Maxwell theory arises as a continuum limit of a lattice gauge theory which does have monopole solutions.\footnote{A more familiar situation in which dynamical monopoles arise are 3$d$ non-abelian gauge theories in which the gauge symmetry is broken to a subgroup containing $U(1)$. Charged matter fields can then act as a localized source of magnetic flux. While we believe some of our results should also apply to this case, it is more difficult to construct non-abelian non-local gauge theories. We therefore only consider the lattice UV completion which leads to a relatively simple non-local theory.}
When doing the functional integral over $A_\mu$, one also must also sum over the monopole/anti-monopole configurations. The monopole gas gives rise to a (non-perturbative) effective screening potential which in some cases leads to a massive photon and a linear potential between charges.

Let us briefly review this mechanism in broad strokes. When $s=2$, 3$d$ electric-magnetic duality allows us to recast Maxwell theory as a theory of a compact scalar $\phi$ whose action is given by
\begin{equation}
	S=-\frac{e^2}{8\pi^2} \int d^3x\, \phi\nabla^2\phi\,.
	\label{E:3dSG}
\end{equation}
By UV completing Maxwell theory on a lattice, Polyakov carries out a resummation of monopole solutions to obtain an effective cosine potential, $V=\lambda\cos(\phi)$, where $\lambda$ is related to $e$ and the lattice cutoff. In the infrared the cosine term dominates the dynamics and Wilson lines can be shown to possess an area law indicating confinement.

It is natural to ask whether a similar confinement mechanism comes into play in the effective description of the boundary dynamics and large $N_f$ QED of a compact $U(1)$ field given by the action \eqref{E:U1action0}. The current work provides an extended analysis of the resulting dynamics and closely follows  \cite{Polyakov:1975rs}. In section \ref{sec:Nonlocal} we discuss the non-local action \eqref{E:U1action0} and its associated dual photon. In section \ref{sec:Instantons} we UV complete \eqref{E:U1action0} on a lattice and sum up monopole solutions to obtain a non-local (generalized) version of the sine-Gordon theory in three dimensions generalizing \eqref{E:3dSG}.

In section \ref{sec:SineGordon} we perform a perturbative RG calculation of this generalized sine-Gordon theory to third-order in the coupling constant of the cosine potential, $\lambda$. Refs. \cite{Kleinert:2002qk,Herbut:2003bs} previously studied the RG flow of this model at orders $\lambda$ and $\lambda^2$ respectively. The first-order beta function for $\lambda$ reveals that the theory exhibits a phase transition with the cosine potential changing from being an irrelevant to a relevant deformation of the generalized free theory as the value of the coupling $e$ is increased, a fact previously noted by \cite{Kleinert:2002qk} in the context of an Abelian Higgs model.  But, as observed in  \cite{Herbut:2003bs}, second-order perturbation theory reveals that generically the RG flow generates a relevant local kinetic term steering the theory towards local, confining Maxwell theory. At third-order in $\lambda$, we find that as the result of a delicate cancellation of terms, the theory remains renormalizable, despite the non-locality of the model invalidating standard proofs of renormalizability.\footnote{Ref. \cite{alves2022supersymmetric} recently provided a proof of all-loop renormalizability of the supersymmetric version of fractional-derivative Maxwell theory \eqref{E:U1action0} with $s=1$.}

When the cosine potential is relevant, the compact scalar acquires a mass, and electric charges become confined. The arguments attesting to these facts in the local case carry over to generalized Maxwell theory, subject to some technical obstacles, which we tackle and resolve in section \ref{sec:Confinement} where we compute the expectation value of a Wilson loop. In particular, through a numerical study of the generalized sine-Gordon equation, we demonstrate the existence of a field configuration in which the compact scalar jumps when crossing the plane bounded by a large loop. This allows us to use the method of steepest descent to evaluate the expectation value of the loop and finally observe area law behavior.

Section \ref{sec:Discussion} wraps up the paper with a discussion of our results and the implications of the RG flow of the compact scalar theory for generalized Maxwell theory in the original gauge field formulation. A possible connection to the `string' mechanism of confinement is briefly discussed.

\section{Non-locality and the dual photon}
\label{sec:Nonlocal}

Our goal is to study the infrared physics of a non-local version of the Polyakov model, which we will show is essentially the generalized free Maxwell action in equation \eqref{E:U1action0} deformed by relevant interactions coming from non-perturbative effects. Because the free theory is expressed in terms of a fractional-derivative operator $D^\sigma=(-\nabla^2)^{\sigma/2}$ with $\sigma \in \mathbb{R}$, we will establish our definition and conventions for this operator. It is particularly simple in momentum space, where it amounts to multiplication by a (possibly fractional) power of the Fourier momentum.

More concretely, let $\mathcal{F}$ denote the operator that implements the Fourier transform on a function, $\mathcal{F}\phi = \tilde{\phi}$. If we let $\pi_\sigma$ be an operator that acts on a function $f(x)$ by multiplying it with a power of the norm of its argument, $\pi_\sigma f(x)=|x|^\sigma f(x)$, then, as in \cite{Huang:2020aao},
the fractional-derivative operator can be expressed as
\begin{align*}
D^\sigma = \mathcal{F}^{-1} \pi_\sigma \mathcal{F}\,.
\end{align*}
This representation makes it clear that the fractional derivative satisfies the composition property, $D^{\sigma_1+\sigma_2}=D^{\sigma_1}D^{\sigma_2}$. 
For general values of $\sigma$, the fractional derivative in position space is an integro-differential operator which acts as \cite{gubser2019non}
\begin{align}
D^\sigma f(x)=
(2\pi)^\sigma\frac{\Gamma(\frac{d+\sigma}{2})}{\pi^{\sigma+d/2}\Gamma(-\frac{\sigma}{2})}\int \frac{d^dy}{|x-y|^{d+\sigma}}
\bigg[
f(y)-\sum_{r=0}^{\lfloor \sigma/2 \rfloor}y^{2r}b_r (\nabla^2)^r f(x)
\bigg]\,,
\end{align}
where the second term is essentially a subtraction to regulate the expression and the coefficients are given by
\begin{align}
b_r=\frac{\Gamma(d/2)}{4^{r}\Gamma(r+d/2)\Gamma(r+1)}\,.
\end{align}
This position space representation makes it clear that the action \eqref{E:U1action0} is non-local,
\begin{align}
\label{E:U1action1}
    S=
    \frac{2^s\Gamma(\frac{1+s}{2})}{16\pi^{3/2}\Gamma(\frac{2-s}{2})e^2}
    \int d^3x\,d^3y\frac{F_{\mu\nu}(x)F^{\mu\nu}(y)}{|x-y|^{1+s}}\,.
\end{align}
In the limit $s\rightarrow 2$, this action tends to the usual local Lagrangian, as can be verified by writing the action as an integral over momentum space.

As mentioned in the introduction, local $U(1)$ Maxwell theory in three dimensions may be written in terms of a dual photon. This is an example of an electric-magnetic duality and is an essential step in incorporating the effects of magnetic monopoles in the full Polyakov model in terms of local field variables. To generalize this construction to the non-local theory \eqref{E:U1action1}, we start with the path integral on $\mathbb{R}^3$ in terms of the electric gauge field:
\begin{align}
	Z & = \int DA \exp\left(-\frac{1}{4 e^2} \int d^3x\, F_{\mu\nu} D^{s-2} F^{\mu\nu} \right) \, .
\end{align}
We take the fluxes $F$ to be in the integer cohomology, $F \in H^2(\mathbb{R}^3; \mathbb{Z})$. $A$ is a connection on this principle $U(1)$ bundle, and, in general may be singular due to the presence of monopole sources. In addition to the $U(1)$ gauge symmetry $A \rightarrow A + d \alpha$, this action also has a global symmetry with conserved current $j = \ast F$ via the Bianchi identity. In modern language, this topological current corresponds to a generalized global $0$-form symmetry \cite{Gaiotto:2014kfa}. Because this theory lacks explicit $A_\mu$ dependence, there is also an additional $1$-form symmetry
\begin{equation}
    A_\mu \rightarrow A_\mu + \Lambda_\mu \, ,
    \label{general1form}
\end{equation}
where $\Lambda$ is an arbitrary flat connection. Physically, local operators such as monopole operators are charged under $0$-form symmetries, while non-local operators (such as Wilson lines) are charged under $1$-form symmetries. In addition to non-local observables of this kind, theories with fractional kinetic terms might also have non-local operators given by $D^\sigma$ acting on a local operator.\footnote{Natural examples of this kind of non-local operator include conserved currents of non-local field theories \cite{Phillips:2019qkc,Heydeman:2020ijz}.} 
While we do not study these non-local operators directly, in principle they have a straightforward definition in the path integral approach similar to the treatment of topological operators. 

The electric-magnetic duality transformation is implemented by integrating in (and then out) certain degrees of freedom, as reviewed, for example, in \cite{wittenlecture}. Abstractly, one takes the action with a global symmetry and couples it to a dynamical gauge field along with a topological term. The intermediate gauge field is then integrated out. We can do this for the 1-form symmetry by first introducing a 2-form gauge field $B_{\mu \nu}$. In addition to the symmetry \eqref{general1form}, we also have
\begin{align}
    B_{\mu \nu} \rightarrow B_{\mu \nu} + \partial_{[\mu}\Lambda_{\nu ]} \, .
\end{align}
To work with an arbitrary gauge parameter $\Lambda$ and not just a flat 1-form, we must covariantize the field strengths with respect to this larger symmetry. An appropriate choice is 
\begin{equation}
    \mathcal{F} = F - B \, .
\end{equation} When $B$ is a flat higher form connection satisfying $dB=0$, this field strength reduces to the usual one by gauging away $B$ (emphasizing that the non-local kinetic term does not modify the local definition of the field strength). By introducing an additional Lagrange multiplier  $\phi$ whose equation of motion enforces the flatness condition, we obtain the path integral
\begin{equation}
    Z = \int \! D A \, D B \, D \phi \,   \exp \left (-\int \! \! d^3x \, \, \frac{1}{4e^2} \mathcal{F}_{\mu \nu} D^{s-2} \mathcal{F}^{\mu \nu} - \frac{i}{4\pi} \int d^3x \, \epsilon^{\mu \nu \rho} \partial_\mu \phi B_{\nu \rho}\right ) \, .
    \label{eq:Z1}
\end{equation}
We have normalized the Lagrange multiplier such that, as a consequence of standard Dirac quantization, the field $\phi$ is a compact field with a periodicity of $2\pi$.

 Integrating out $\phi$ returns us to the original theory.\footnote{We have ignored the possibility that $B$ could have nontrivial holonomies on more general manifolds, but a more careful argument shows $B$ reduces to the trivial flat connection.} Alternately, we can try to eliminate $A$ and then do the path integral over B. To do this, we use the $1$-form gauge symmetry to set $A$ to be zero everywhere.\footnote{While we do not elaborate on it here, the quantization of non-local abelian gauge theories in the Faddeev-Popov or BRST formalism will typically require non-local ghosts. This is the case for a covariant gauge choice such as $\partial_\mu A^\mu = 0$ or the Feynman $\xi$ gauge in the original gauge theory. In the present case, the integral over the ghosts and intermediate gauge fields produces a ratio of determinants which appears as an overall multiplicative factor we do not include explicitly.} The path integral over $A$ is straightforward, 
 and we are left with
\begin{equation}
    Z = \int \! D B \, D \phi \,   \exp \left (-\int \! \! d^3x \, \, \frac{1}{4e^2} B_{\mu \nu} D^{s-2} B^{\mu \nu} - \frac{i}{4\pi} \int d^3x \, \epsilon^{\mu \nu \rho} \partial_\mu \phi B_{\nu \rho}\right ) \, .
\end{equation}
 This action is quadratic in $B$, but in contrast to the local ($s=2$) case, it contains a non-local differential operator acting on $B$. Nevertheless, we can integrate out $B$ by completing the square and solving the classical equation of motion. Using the composition law $D^{\sigma_1+\sigma_2}=D^{\sigma_1}D^{\sigma_2}$ allows us to invert the non-local derivative, leading to the stationary phase configurations
\begin{equation}
	B_{\mu\nu} = -\frac{ie^2}{2\pi}\epsilon_{\mu\nu\rho} D^{2-s} \partial^{\rho} \phi\, .
	\label{E:fielddualstrength}
\end{equation}
In the local ($s=2$) theory, this equation may be used to eliminate $B$ in favor of $\phi$, leading to the dual action \eqref{E:3dSG}, and we assume the same is true for the non-local equation of motion. 
The final result after integrating out the auxiliary potential is:
\begin{equation}
\label{E:dualphotonaction}
	Z = \int D\phi \exp\left( -\frac{e^2}{8\pi^2} \int d^3x\, \partial_{\rho} \phi D^{2-s} \partial^{\rho} \phi \right)\, .
\end{equation}
In the ordinary local case and here, the gauge field has been exchanged with a compact boson and the coupling $e$ has been inverted. The new feature of the non-local case is that the non-local exponent $s$ has also been inverted, in the sense that a higher (lower) derivative gauge theory is exchanged with a lower (higher) derivative scalar theory, respectively. 

Local and non-local operator insertions may also be mapped across the duality transformation, and of particular interest are Wilson lines (discussed in section \ref{sec:Confinement}) and monopole operators which create field configurations satisfying \eqref{E:monopoleflux}.
For $s=1$, the dual action has a logarithmic propagator, and the operator $e^{i\phi}$ becomes a candidate monopole operator. A related observation was made previously in \cite{kapustin1999mirror}, which pointed out that in the large $N_f$ limit of $\mathcal{N}=2$ and $\mathcal{N}=4$ SQED, the exponentiated generalized free field has the correct scaling dimension and vortex charge for a monopole operator but does not, however, carry the appropriate $R$-charge. In the large $N_f$ limit of regular QED$_3$ without supersymmetry, as pointed out in \cite{Borokhov:2002ib}, the scaling dimension of the exponentiated generalized free field no longer matches the monopole operator, a mismatch \cite{Borokhov:2002ib} ascribed to generalized free theory not taking into account fermionic zero modes.\footnote{%
The dimension of the exponentiated generalized free field scales as $N_f/4$ at large $N_f$, whereas Ref. \cite{Borokhov:2002ib} determined the monopole scaling dimension, $\Delta_{\text{monopole}}$, to be given to leading order in $N_f$ by
\begin{align*}
\frac{\Delta_{\text{monopole}}}{N_f}
=\frac{1}{4}
+
\frac{3}{8\,\Gamma(\frac{5}{2})}\sum_{n=2}^\infty
\frac{(-1)^n(n-1)\Gamma(\frac{1}{2}+n)}{\Gamma(n+3)}\zeta(n)
= 0.2650955...\,.
\end{align*}
For monopole operators with higher vorticity, the mismatch gets increasingly worse.} This issue is not present for the purely bosonic action \eqref{E:U1action0}, nor the issue of higher order interaction terms generated by integrating out fermions, so that in our setting we can confidently equate $e^{i\phi}$ with the monopole operator. We corroborate this identification in the next section by showing how a sum over instanton configurations gives rise to the proliferation of this operator.

\section{Monopole solutions and proliferation}
\label{sec:Instantons}

So far, we have reviewed the non-local Maxwell action and the equivalence with a compact scalar in three dimensions. At the level of the field theory, we do not yet have a mechanism to supply monopole solutions. This can be incorporated by adding additional UV degrees of freedom which flow to the Maxwell theory (or the non-local variant) in the infrared. Instead of considering for instance a non-abelian gauge theory at a point in the moduli space with an unbroken $U(1)$, we will instead follow Polyakov's original proposal of UV completion on a lattice. Suppose that the action \eqref{E:U1action1} is the low energy limit of an action $S_{UV}$ which satisfies 
\begin{equation}
\label{E:Speriodicity}
    S_{UV}(F_{\mu\nu}) = S_{UV}(F_{\mu\nu} + 2\pi N_{\mu\nu})
\end{equation} 
where $N_{\mu\nu}$ is a lattice-valued tensor field.  
It is possible to find a lattice action that satisfies \eqref{E:Speriodicity} and flows to \eqref{E:U1action1} in the infrared. 
Indeed, consider a three-dimensional cubic lattice with lattice spacing $a$, link variables
\begin{equation}
    U_{\mu}(x) = e^{-i a A_{\mu}(x)}\,,
\end{equation}
and plaquette variables
\begin{align}
\begin{split}
    W_{\mu,\nu}(x)
    &=U_\mu(x)U_\nu(x+\hat{\mu})
    U_\mu^\dagger(x+\hat{\nu})
    U_\nu^\dagger(x)
    \\
    %&= 1- ia^{2}F_{\mu\nu}(x) -\frac{a^{4}}{2}F_{\mu\nu}(x)^2 +\mathcal{O}(a^6)\,,
    &= e^{-i a^2 F_{\mu\nu}}
\end{split}
\end{align}
(with $F$ the field strength associated with $A_{\mu}$).
The non-local action \eqref{E:U1action1} is given by the continuum limit of
\begin{equation}
\label{E:SUV}
    S_{UV} = -\frac{2^s\,\Gamma(\frac{1+s}{2}) a^2}{16\pi^{3/2}\, e^2\,\Gamma(\frac{2-s}{2})} \sum_{\mu\neq\nu}\sum_{x,y}W_{\mu,\nu}(x) \frac{W_{\mu,\nu}(y) - 2}{|x-y|^{1+s}}\,,
\end{equation}
up to a constant term.

Monopole solutions located at $x^{\mu}_i$ with charges $q_i$, attached to a Dirac string oriented in, for instance, the $x^3$ direction,
\begin{equation}
\label{E:Monopolestring}
     \frac{1}{2}\epsilon^{\mu\nu\rho}F_{\nu\rho} = \sum_i \frac{q_i}{2}\frac{(x-x_i)^{\mu}}{|x-x_i|^3} - \sum_i 2\pi q_i \delta^{\mu}_{3} \theta(x^3-x_{i}^{3})\delta(x^1-x_{i}^{1})\delta(x^2-x_{i}^{2})
\end{equation}
are solutions to the Bianchi identity, 
\begin{equation}
        \epsilon^{\mu\nu\rho}\partial_{\mu}F_{\nu\rho}=0\,,
\end{equation}
and also solutions to the free, local, Maxwell equations (in a distributional sense).
We will refer to the first term on the right hand side of \eqref{E:Monopolestring} as the monopole contribution to the field configuration and to the second term on the right hand side of \eqref{E:Monopolestring} as the string contribution to it. 
Due to \eqref{E:Speriodicity}, the string contribution will not contribute to the action $S_{UV}$. 

Any solution to the local version of the Maxwell equations is also a solution to the non-local equations of motion, c.f., the discussion surrounding \eqref{nonlocalFourier}.  Solutions to \eqref{nonlocalFourier} which are not solutions to the local Maxwell equations are distributions (in momentum space) with support at large $k^{\mu}$. Their Fourier transform would correspond to a highly oscillatory function which would vanish when integrated over a reasonable test function from the stationary phase approximation.
Since the monopole contribution to the field strength solves the local Maxwell equation, we must take it into account when evaluating a saddle point related to the infrared theory associated with \eqref{E:SUV}; this was one of the insights in \cite{Polyakov:1975rs}.

Inserting the monopole contribution to the field configuration \eqref{E:Monopolestring} into the non-local action \eqref{E:U1action1} we find, by making repeated use of
\begin{equation}
    \int \frac{d^dx}{|x-a|^{\alpha}|x-b|^{\beta}} = 
    \pi^{d/2}\frac{\Gamma(\frac{d-\alpha}{2})\Gamma(\frac{d-\beta}{2})\Gamma(\frac{\alpha+\beta-d}{2})}
    {\,\Gamma(\frac{\alpha}{2})\Gamma(\frac{\beta}{2})\Gamma(\frac{2d-\alpha-\beta}{2})}|a-b|^{d-\alpha-\beta}\,,
\end{equation}
that, in the $d\to3$ limit, 
\begin{equation}
\label{E:Smonopole}
    S = \frac{K}{e^2} \sum_{i\neq j} \frac{q_i q_j}{|x_i-x_j|^{s-1}}
    +
    \frac{\epsilon}{e^2} \sum_i q_i^2\,,
\end{equation}
where
\begin{equation}
\label{E:Kappa}
    K = \frac{2^s\pi^{\frac{1}{2}}\Gamma(\frac{s-1}{2})}{8\,\Gamma(\frac{4-s}{2})} 
    \,,
    \hspace{15mm}
    \epsilon = \frac{K'}{a^{s-1}}\,,
\end{equation}
and $K'$ is a scheme dependent dimensionless number. When $s=1$, $\epsilon$ becomes dimensionless and \eqref{E:Smonopole} takes the logarithmic form of a Coulomb gas.

A saddle point approximation of the monopole contributions to the Euclidean generating function is given by \begin{equation}
\label{E:monopole}
    Z \sim \sum_{N=0}^{\infty} \frac{1}{N!} \left[\prod_{i=1}^{N} \, \, \sum_{q_i=\pm 1,2,\dots} \! \! a^{-3}\int d^3 x_i \right] e^{-S}\,,
\end{equation}
with $S$ given in \eqref{E:Smonopole}. 
Using 
\begin{multline}
    \int D\varphi \exp\left( -\frac{K}{e^2}\frac{16\pi^{3/2}\Gamma(2-\frac{s}{2})}{2^s\Gamma(\frac{s-1}{2})} \frac{1}{C^2} \int d^3x \left( \varphi D^{4-s} \varphi + 2 i C \sum_i \varphi q_i \delta(x-x_i)\right) \right)
    \\
    \propto
    \exp\left( 
    -\frac{K}{e^2} \sum_{i,j} \int d^3x d^3y \frac{q_i q_j \delta(x-x_i)\delta(y-x_j)}{|x-y|^{s-1}}
    \right)\,,
\end{multline}
with $C$ an arbitrary constant,
we find
\begin{align}
\begin{split}
\label{E:MSum}
    Z &\sim 
    \int D\varphi\, e^{-\frac{4\pi^2}{e^2C^2} \int d^3x\, \frac{1}{2}\varphi D^{4-s} \varphi} \sum_{N=0}^{\infty}  \frac{1}{N!} \prod_{i=1}^N  \frac{1}{a^3}\int d^3x_i\, \sum_{n=1}^\infty e^{-\frac{n^2 \epsilon}{e^2}}2\cos\left(n\frac{4\pi^2\varphi}{e^2C}\right)  \\
    &=  \int D\varphi\, e^{- S_{\varphi}}\,,
\end{split}
\end{align}
where
\begin{equation}
\label{E:Sphi}
    S_{\varphi} = \frac{4\pi^2}{e^2 C^2} \int d^3x \, \bigg(\frac{1}{2} \varphi D^{4-s} \varphi - \frac{e^2 C^2 }{2\pi^2a^3}\sum_{n=1}^\infty e^{-\frac{n^2 \epsilon}{e^2}}
    \cos\Big(n\frac{4\pi^2 \varphi}{ e^2 C}\Big)\bigg)\,.
\end{equation}

After a field redefinition to a new field $\phi=4\pi^2\varphi/(e^2C)$, we may write \eqref{E:Sphi} in the form 
\begin{equation}
\label{E:NLSG1}
    S = \int d^3x \bigg(\frac{1}{2g}\phi D^\sigma \phi - \sum_{n=1}^\infty\lambda_n \cos(n\phi)\bigg),
\end{equation}
with $\sigma=4-s$, $g=4\pi^2/e^2$, and $\lambda_n = \frac{2}{a^3}e^{-\frac{\epsilon n^2}{e^2}}$. We observe that the kinetic term for $\phi$ that we have now obtained through an instanton summation exactly agrees with the dual photon action \eqref{E:dualphotonaction} that was derived from a duality transformation of the free gauge theory. Having in each case normalized $\phi$ to be $2\pi$-periodic, the coefficients of the kinetic terms match.

\section{RG flow of non-local sine-Gordon theory}
\label{sec:SineGordon}

Equation \eqref{E:NLSG1} is a non-local version of the sine-Gordon action with higher harmonics present. Because the bare values of $\lambda_n$ decay exponentially with $n^2$, and because $\cos(n\phi)$ will turn out to be irrelevant for $g>12\pi^2/n^2$, we will retain only the leading harmonic in the following and rename $\lambda_1$ as simply  $\lambda$. Since fields associated with non-local kinetic terms of the type appearing in \eqref{E:NLSG1} (for $s$ not even) do not receive wavefunction renormalization, (see e.g., \cite{Heydeman:2020ijz} for a gauge theory version of this non-renormalization theorem), one might expect $\lambda$ to be the only running parameter in the theory once we have discarded the $\lambda_n$ with $n>1$. However, if $s$ is sufficiently large, a local kinetic term will become relevant and might dominate the dynamics. This was not an issue in the free theory, and also in the original Maxwell theory description in which the local term was irrelevant. In general however, to understand the renormalization group flow associated with \eqref{E:NLSG}, we consider the action
\begin{equation}
\label{E:ourS}
    S = \int d^3x\bigg( \frac{1}{2g}\phi D^\sigma \phi - \frac{h}{2} \phi \nabla^2 \phi  - \lambda \cos(\phi)\bigg)\,,
\end{equation}
where we choose $h$ to be positive. Expanding the cosine shows that $\lambda$ controls the mass of the dual photon, so the flow of this parameter controls whether or not the theory becomes confining in the infrared due to monopole proliferation.

As stated earlier, the beta function for the non-local kinetic term, $\beta_g$ will not receive quantum corrections for non even $s$ and will vanish for $\sigma=3$ ($s=1$), which is the case we will focus on.
To compute the beta functions for $h$ and $\lambda$, $\beta_h$ and $\beta_{\lambda}$, we will compute the one-particle irreducible (1-PI) 2-point vertex, $\Sigma_2$, and renormalize the couplings, $h$ and $\lambda$, so that the full propagator $G$, given by
\begin{equation}
\label{E:fullG}
	G^{-1} = G_{f}^{-1} - \Sigma_2
\end{equation}
with $G_f$ the propagator in the free theory, is finite.   We will use a hard IR cutoff $\mu$ in momentum space, and a hard UV cutoff $a$ in real space, the latter in line with our lattice construction from the previous section. 
On a practical level, we introduce dimensionless bare couplings by setting
\begin{equation}
	g=g_0\,,
	\qquad
	\lambda = \frac{\lambda_0}{a^3}\,,
	\qquad
	h = \frac{h_0}{a}\,.
\end{equation}
In the following we will express $h_0$ and $\lambda_0$ in terms of renormalized couplings $h_r$ and $\lambda_r$, chosen such as to ensure that $G$ does not diverge in the $a\to 0$ limit \cite{AGG}. 
We work perturbatively in $\lambda_0$. 
Note that since $\lambda \to -\lambda$ together with $\phi \to \phi+\pi$ is a symmetry of the action, $h$ can get perturbatively renormalized only at even orders in $\lambda$, and $\lambda$ itself can only get renormalized at odd powers.

Evaluating certain loop integrals below requires the free propagator in position space, given by
\begin{align}
\label{E:Gfree}
	G_{f}(x) &= \int_{\mu}^{\infty} k^2 dk \int d\Omega_k \frac{e^{2\pi i k \cdot x}}{\frac{1}{g}(2\pi k)^3 + \frac{h_0}{a}(2\pi k)^2}  \\ \nonumber
			  & = \frac{g}{2\pi^2} \int_{2\pi x \mu}^{\infty} \frac{\sin(\rho)}{\rho^2 + \frac{g h_0 x}{a} \rho} d\rho \\ \nonumber
			  &=\frac{g }{4 \pi ^2 \chi }\Big(2 \cos (\chi ) \text{Si}(2 \pi  x \mu +\chi )-2 (\sin (\chi ) \text{Ci}(2 \pi  x \mu +\chi )+\text{Si}(2 \pi  x \mu ))-\pi  \cos (\chi )+\pi \Big)\Bigg|_{\chi = \frac{g h_0 x}{a}} \,,
\end{align}
where $\hbox{Si}$ and $\hbox{Ci}$ are the sine integral and cosine integral functions respectively,
\begin{equation}
	\hbox{Si}(y) = \int_0^y \frac{\sin(z)}{z} dz
	\qquad
	 \hbox{Ci}(y) = -\int_y^\infty \frac{\cos(z)}{z} dz\,.
\end{equation}
Expanding $G_f$ around small values of $x$ we find
\begin{equation}
	G_f(x) = -\frac{g}{2\pi^2} \ln\left(\left(\frac{g h_0}{a} + 2\pi \mu\right) c x\right) + \frac{g^2 h_0 x}{8 \pi a}+ \mathcal{O}(x^2)\,,
\end{equation}
where $c= e^{\gamma-1}$ with $\gamma$ Euler's constant. Note that if $h_0=0$, then the IR cutoff is needed to make $G_f(x)$ well-defined. Since we are assuming a local term has been generated, we  may set $\mu$ to 0 in almost all of our computations. We will comment on the role of the IR regulator when relevant.

In order for $G_{f}$ to be finite at tree-level, $h_0$ and $h_r$ must be related via
\begin{equation}
\label{E:hr1}
	h_0 = \kappa a h_r + \mathcal{O}(\lambda_r)\,,
\end{equation}
where $\kappa$ is the RG scale. To compute the 1-PI 2 point vertex we work perturbatively in $\lambda_0$, decomposing $\Sigma_2$ as 
\begin{equation}
	\Sigma_2 = \sum_{n=1} \Sigma_2^{(n)} \,,
\end{equation}
where $\Sigma_2^{(n)}$ is the term of order $(\lambda_0)^n$ in the expansion of $\Sigma_2$ with respect to $\lambda_0$. In the remainder of this section we will compute the $\Sigma_2^{(n)}$ up to order $n=3$ and use this result to compute the $\beta$ functions associated with $h_r$ and $\lambda_r$. Our results extend those in \cite{Kleinert:2002qk,Herbut:2003bs}, first found in a somewhat different context.

\subsection{Order $\lambda_0$}

The leading order in $\lambda_0$ correction to the propagator comes from the cosine term. The vertex involves a sum of bubble diagrams, and we find:
\begin{align}
\begin{split}
\\[-40pt]
	\Sigma_2^{(1)} &
	=
	\hspace{3mm}
	\begin{tikzpicture}[baseline=0 cm]
	\begin{feynman}
	\vertex (ll) at (-0.7,0) {$p$};
	\vertex (r) at (0.3,0);
    \filldraw[color=black] (0,0) circle (1pt);
	\vertex (l) at (-0.3,0);
	\diagram* {
	(l) --  (r) ,
	};
	\end{feynman}
	\end{tikzpicture}\,
		\hspace{3mm}
	+
				\hspace{-3mm}
	\begin{tikzpicture}[baseline=0 cm]
	\begin{feynman}
		\vertex (ll) at (-0.7,0) {$p$};
	\vertex (c) at (0,0);
	\vertex (r) at (0.4,0);
	\vertex (l) at (-0.4,0);
	\diagram* {
	(l) -- (c) -- (r) ,
	(c) -- [out=135, in=45, loop, min distance=2 cm] c,
%	(c) -- [out=135, in=45, loop, min distance=2 cm] c,
	};
	\end{feynman}
	\end{tikzpicture}\,
			\hspace{-5mm}
	+
			\hspace{-5mm}
	\begin{tikzpicture}[baseline=0 cm]
	\begin{feynman}
	\vertex (ll) at (-0.85,0) {$p$};
	\vertex (c) at (0,0);
	\vertex (r) at (0.5,0);
	\vertex (l) at (-0.5,0);
	\diagram* {
	(l) -- (c) -- (r) ,
	(c) -- [out=160, in=90, loop, min distance=2 cm] c,
	(c) -- [out=90, in=20, loop, min distance=2 cm] c,
	};
	\end{feynman}
	\end{tikzpicture}\,
				\hspace{-5mm}
	+
				\hspace{5mm}
	\ldots \\[8pt]
	&= \frac{\lambda_0}{a^3} \sum_{n=0}^\infty (-1)^n \frac{(2n-1)!!}{(2n)!} G_f(0)^n\\
	&= \frac{\lambda_0}{a^3}e^{-\frac{1}{2}G_f(0)}\,.
 \end{split}
\end{align}
Unsurprisingly, the expression for $\Sigma_2^{(1)}$ in terms of $G_f$ is identical to the one obtained in the local sine-Gordon theory, c.f., \cite{AGG}. The difference between the expression for $\Sigma_2^{(1)}$ in the local and non-local theories is encoded in the kinetic term $G_f$.

Using \eqref{E:Gfree} and `$a$' as a hard UV cutoff, we find 
\begin{equation}
\label{E:Sigma21}
\Sigma_2^{(1)} = 
	\lambda_0\, a^{\frac{g}{4 \pi ^2}-3} (c g h_r\kappa)^{\frac{g}{4 \pi ^2}} \left(1 - \frac{g^2 h_r}{16 \pi }\kappa a + \mathcal{O}(a^2) \right)\,.
\end{equation}
When $g>12\pi^2$ the expression for $G^{-1}_{f}-\Sigma_2^{(1)}$ is finite. For $g\leq 12\pi^2$ we must have
\begin{equation}
\label{E:rorder1}
	\lambda_0 = \lambda_r (\kappa a)^{3-\frac{g}{4\pi^2}} + \mathcal{O}(\lambda_r^3)\,
\end{equation}
in addition to \eqref{E:hr1} in order for the propagator to be divergenceless.
Note that, similar to the local sine-Gordon theory, the tree level scaling dimension of $\lambda$ is $3-g/(4\pi)^2$ due to the non perturbative contributions coming from $g$. For $g>12\pi^2$, we see that the cosine interaction is irrelevant. For this reason, we focus on the regime $g\leq 12\pi^2$ in the remainder of this section.

\subsection{Order $\lambda_0^2$}
We now compute the next order of corrections via the diagrams:
\begin{align}
\begin{split}
	\begin{tikzpicture}[baseline=-0.1 cm]
	\begin{feynman}
	\vertex (r) at (1,0);
	\vertex (l) at (-1,0);
	\vertex (fl) at (-1.5,0) {$p$};
	\vertex (fr) at (1.3,0);
	\vertex (d1) at (0,0.3) {};
	\vertex (d2) at (0,0.15) {};
	\vertex (d3) at (0,0.0) {};
	\diagram* {
	(fl) [particle=\(p\)]--  (l),
	(r) -- (fr),
	(l) --[out=90, in=90, loop, min distance=1.7 cm, edge label=\(k_1\)] (r),
	(l) --[out=70, in=110, loop, min distance=1 cm, edge label=\(k_2\)] (r),	
	(l) --[out=270, in=270, loop, min distance=1.7 cm, edge label=\(k_n\)] (r),
	(l) --[out=290, in=250, loop, min distance=1 cm, edge label=\(k_{n-1}\)] (r),	
	};
	\node at (d1)[circle,fill,inner sep=0.7pt]{};
	\node at (d2)[circle,fill,inner sep=0.7pt]{};
	\node at (d3)[circle,fill,inner sep=0.7pt]{};
	\end{feynman}
	\end{tikzpicture}
	\quad
	&=
	\frac{(n+1)^2 n! }{((n+1)!)^2} \left(\prod_{i=1}^{n} \int d^3k_i\, \hat{G}(k_i) \right) \delta\left(\sum_{i=1}^n k_i + p\right) \\
	&=
	\frac{1}{n!} \int d^3x e^{2\pi x \cdot p} G(x)^n\,,
\end{split}
\end{align}
and
\begin{align}
\begin{split}
	\begin{tikzpicture}[baseline=0.8 cm]
	\begin{feynman}
	\vertex (r) at (0.5,0);
	\vertex (l) at (-0.7,0)  {$p$};
	\vertex (c) at (0,0) ;
	\vertex (t) at (0,2);
	\vertex (d1) at (-0.45,1) {};
	\vertex (d2) at (-0.3,1) {};
	\vertex (d3) at (-0.15,1) {};
	\diagram* {
	(l) --  (r),
	(c) --[out=160, in=200, loop, min distance=2.5 cm, edge label=\(k_1\)] (t),
	(c) --[out=140, in=220, loop, min distance=1.7 cm, edge label=\(k_2\)] (t),	
	(c) --[out=40, in=-40, loop, min distance=1.7 cm, edge label=\(k_{n-1}\)] (t),	
	(c) --[out=20, in=-20, loop, min distance=2.5 cm, edge label=\(k_n\)] (t),
	};
	\node at (d1)[circle,fill,inner sep=0.7pt]{};
	\node at (d2)[circle,fill,inner sep=0.7pt]{};
	\node at (d3)[circle,fill,inner sep=0.7pt]{};
	\end{feynman}
	\end{tikzpicture}
	\quad
	&=
	-\frac{(n+2)!}{ n!  (n+2)!} \left(\prod_{i=1}^{n} \int d^3k_i\, \hat{G}(k_i) \right) \delta\left(\sum_{i=1}^n k_i \right) \\
	&=
	-\frac{1}{n!} \int d^3x G(x)^n\,.
\end{split}
\end{align}
We find
\begin{align}
\begin{split}
\label{E:Sigma22}
	\Sigma_2^{(2)}(p) &= \left(\Sigma_2^{(1)}\right)^2 \left( \sum_{n=1}^{\infty}
		\begin{tikzpicture}[baseline=-0.1 cm]
		\begin{feynman}
			\vertex (r) at (1,0);
			\vertex (l) at (-1,0);
			\vertex (fl) at (-1.5,0) {};
			\vertex (fr) at (1.3,0);
			\vertex (d1) at (0,0.3) {};
			\vertex (d2) at (0,0.15) {};
			\vertex (d3) at (0,0.0) {};
			\diagram* {
			(fl) [particle=\(p\)]--  (l),
			(r) -- (fr),
			(l) --[out=90, in=90, loop, min distance=1.7 cm, edge label=\(k_1\)] (r),
			(l) --[out=70, in=110, loop, min distance=1 cm, edge label=\(k_2\)] (r),	
			(l) --[out=270, in=270, loop, min distance=1.7 cm, edge label=\(k_{2n+1}\)] (r),
			(l) --[out=290, in=250, loop, min distance=1 cm, edge label=\(k_{2n}\)] (r),	
			};
			\node at (d1)[circle,fill,inner sep=0.7pt]{};
			\node at (d2)[circle,fill,inner sep=0.7pt]{};
			\node at (d3)[circle,fill,inner sep=0.7pt]{};
		\end{feynman}
		\end{tikzpicture}
	+ \sum_{n=1}^{\infty}
		\begin{tikzpicture}[baseline=0.8 cm]
		\begin{feynman}
			\vertex (r) at (0.5,0);
			\vertex (l) at (-0.7,0)  {$p$};
			\vertex (c) at (0,0) ;
			\vertex (t) at (0,2);
			\vertex (d1) at (-0.6,1) {};
			\vertex (d2) at (-0.45,1) {};
			\vertex (d3) at (-0.3,1) {};
			\diagram* {
			(l) --  (r),
			(c) --[out=160, in=200, loop, min distance=2.5 cm, edge label=\(k_1\)] (t),
			(c) --[out=140, in=220, loop, min distance=1.7 cm, edge label=\(k_2\)] (t),	
			(c) --[out=40, in=-40, loop, min distance=1.7 cm, edge label=\(k_{2n-1}\)] (t),	
			(c) --[out=20, in=-20, loop, min distance=2.5 cm, edge label=\(k_{2n}\)] (t),
			};
			\node at (d1)[circle,fill,inner sep=0.7pt]{};
			\node at (d2)[circle,fill,inner sep=0.7pt]{};
			\node at (d3)[circle,fill,inner sep=0.7pt]{};
		\end{feynman}
		\end{tikzpicture}
	\right) \\
	&=
	4\pi (\Sigma_2^{(1)})^2 \int_a^{\infty} dx\, x^2 \bigg(\frac{\sin(2\pi p x)}{2\pi p x} \Big(\sinh\big(G_f(x)\big) - G_f(x)\Big) + 1-\cosh\big(G_f(x)\big) \bigg)\,.
\end{split}
\end{align}
Once again, this expression is identical in form to the order $\lambda_0^2$ correction of the local sine-Gordon theory which can be found in, e.g., \cite{AGG}.

We are interested in terms in $\Sigma_2^{(2)}$ which are divergent in the limit $a\to 0$. Such terms will arise from the lower end of the integral. Expanding the sine in powers of $p$, we find, to leading order
\begin{equation}
\label{E:p0term}
	\Sigma_2^{(2)}(0) =
	4\pi (\Sigma_2^{(1)})^2 \int_a^{\infty} dx\, x^2 \left( -e^{-G_f(x)} - G_f(x) + 1 \right)\,.
\end{equation}
Given the expansion in the last line of \eqref{E:Gfree} we see that the integrand is such that the integral in \eqref{E:p0term} remains finite in the $a\to 0$ limit.

Next we have
\begin{align}
\begin{split}
\label{E:Sigma22}
	\Sigma_2^{(2)}(p) - \Sigma_2^{(2)}(0) & = \sum_{n=1}^\infty (-1)^n \left(\Sigma_2^{(1)}\right)^2 p^{2n} \frac{\pi^{2n+1} 2^{2n+2}}{(2n+1)!} \int_a^{\infty} dx\, x^{2n+2} \Big(\sinh(G_f) - G_f\Big) \\
	&= \sum_{n=1}^\infty (-1)^n \left(\Sigma_2^{(1)}\right)^2 p^{2n} \frac{\pi^{2n+1} 2^{2n+1}}{(2n+1)!} \left( \int_a^{\infty} dx\, x^{2n+2} e^{G_f(x)} + \mathcal{O}(a^{2n+3}) \right).
\end{split}
\end{align}
(where, in the first equality, we assumed that we can swap the summation over momentum modes with an integral over $x$).

It is convenient to define
\begin{equation}
	I_k(a) = \int_a^{\infty} dx\, x^{2k+2} e^{G_f(x)} \,,
\end{equation}
The dependence of $I_1$ on $a$ in the small $a$ limit will provide us with the information we need on the renormalization of $h_0$ at this order in perturbation theory. If $I_k$ diverges for $k>1$, then this implies that we need to add higher order local kinetic terms to the action. As is the case when studying local theories, these higher order kinetic terms are irrelevant so we may safely ignore them. 
Omitting constant terms, we find
\begin{multline}
\label{E:Ik}
	I_k(a) = \left(  g\kappa h_r c \right)^{-\frac{g}{2\pi^2}}
	\\
	\times 
	\begin{cases} 
		 a^{2k+3-\frac{g}{2\pi^2}} \left(\frac{1}{\frac{g}{2\pi^2}-\left(2k+3\right)}+ \frac{g^2 h_r }{8 \pi}\frac{\kappa a}{\frac{g}{2\pi^2}-\left(2k+4\right)}+ \mathcal{O}(a^2) \right)& \substack{g \neq (2k+n)2\pi^2  \\  n = 3,\,4,\,5,\,...} \\
		- \ln(a) - \frac{g^2 h_r}{8 \pi} \kappa a  + \mathcal{O}(a^2) & g = (2k+3)2\pi^2 \\
		\frac{1}{a} \left(1-  \frac{g^2 h_r}{8 \pi} \kappa a \ln(a) +  \mathcal{O}(a^2) \right) & g = (2k+4)2\pi^2  \\
		\ldots & \ldots
	\end{cases}
\end{multline}
Inserting \eqref{E:Sigma21} and \eqref{E:Ik} into \eqref{E:Sigma22}, we find
\begin{align}
\begin{split}
\label{E:Sigma22e}
	\frac{\partial}{\partial p^2} \Sigma_2^{(2)}\Big|_{p=0} =&
	 \lambda_r^2 \begin{cases}
	-\frac{2\pi^5}{3} (a\kappa)^{5-\frac{g}{2\pi^2}}\kappa \left(\frac{4}{g-10\pi^2} + \frac{g^2 h_r \pi}{(g-10\pi^2)(g-12\pi^2)} (a\kappa) + \mathcal{O}(a^2)\right)
		& \substack{g \neq 2n \pi^2 \\ n=5,\,6,\,\ldots} \\
	\frac{4}{3}\pi^3 \kappa \ln(a) + \frac{50 h_r \pi^6 \kappa}{3} \kappa a\left(1-\ln(a)\right) + \mathcal{O}(a^2) & g = 10 \pi^2 \\
	-\frac{4 \pi^3}{3 a} +24 \pi^6 g^2 h_r \kappa \left(1 +  \ln a \right) + \mathcal{O}(a) & g =12 \pi^2 \\
	\ldots & \ldots
	\end{cases}
\end{split}
\end{align}
For $g<10\pi^2$ we see that $\frac{\partial}{\partial p^2} \Sigma_2^{(2)}$ is finite in the $a\to 0$ limit. For $g=2n\pi^2$ with $n=5,\,6,\,\ldots$, $\frac{\partial}{\partial p^2} \Sigma_2^{(2)}\Big|_{p=0} $ will possess logarithmic divergences in the $a\to 0$ limit. As $g$ increases beyond $g=10\pi^2$, more terms in the expansion of \eqref{E:Sigma22e} will contribute to the ultraviolet divergences of $\frac{\partial}{\partial p^2} \Sigma_2^{(2)}$. 

The divergences in the Green's function \eqref{E:fullG} arising from the $\mathcal{O}(p^2)$ terms in $\Sigma_2^{(2)}$ may be removed by an order $\lambda_r^2$ correction to $h_r$. More explicitly, in order for the order $p^2$ term in $\frac{h_0}{a} (2\pi p)^2 - \Sigma_2^{(2)}(p)$ to remain finite in the $a\to0$ limit we may use a prescription of the form
\begin{equation}
\label{E:h0order2}
	\frac{h_0}{a\kappa} = h_r 
	+ \lambda_r^2 
	\begin{cases}
		0 & 0<g < 10 \pi^2 \\
		\frac{\pi}{3} \ln(\kappa a) + C_3 h_r & g = 10 \pi^2 \\
		-\frac{2\pi^3}{3(g-10\pi^2)}(a\kappa)^{5-\frac{g}{2\pi^2}} +C_1(g)+C_2(g) h_r & 10\pi^2 < g < 12 \pi^2 \\
		-\frac{\pi}{3 a \kappa} + 6 \pi^4 h_r \ln(\kappa a)	 + C_4 & g=12\pi^2 \\
		\ldots & \ldots \\
		\end{cases}\,,
\end{equation} 
up to order $\lambda_r^4$ corrections. Here $C_1(g)$, $C_2(g)$, $C_3$ and $C_4$ are prescription dependent finite counterterms which we will choose later. Other counterterms are also allowed but we have not included them since, as we will see later, they will not affect the beta function for $h_r$.

\subsection{Order $\lambda_0^3$}
At order $\lambda_0^3$ there are eight classes of diagrams that contribute to $\Sigma_2^{(3)}$. As explained earlier, since $\lambda \to -\lambda$ and $\phi \to \phi+\pi$ is a symmetry, $h_r$ will not be renormalized  at order $\lambda_0^3$. So we may evaluate the diagrams at zero momentum. Since this computation is identical to that of the local theory, we will not write out the diagrams explicitly, but quote the result of \cite{AGG},
\begin{equation}
\label{E:Sigma23}
	\Sigma_2^{(3)} = \left(\Sigma_2^{(1)}\right)^3 \frac{1}{8} \int_a^{\infty} d^3x \int_a^{\infty} d^3y \,\bigg( e^{G_f(x)+G_f(y)-G_f(x-y)}-e^{G_f(x)} - e^{G_f(y)}\bigg)\,.
\end{equation}
Since the dependence of $\Sigma_2^{(3)}$ on the UV cutoff is tied to the lower end of the integration region in \eqref{E:Sigma23}, we separate the integration into four regions:
\begin{align}
\begin{split}
	\bf{I}_1 &= \{ x,y \,| \,a< |x| < \Delta,\, a< |y| < \Delta \} \\
	\bf{I}_2 &= \{ x,y \,| \,a< |x| < \Delta,\, \Delta< |y| < \infty \} \\
	\bf{I}_3 &= \{ x,y \,| \,\Delta < |x| < \infty,\, a < |y| < \Delta \} \\	
	\bf{I}_4 &= \{ x,y \,| \,\Delta < |x| < \infty,\, \Delta< |y| < \infty \}\,,
\end{split}
\end{align}
where $\Delta$ is a length scale much bigger than $a$. Because we are only interested in the part of $\Sigma_2^{(3)}$ that diverges in the $a\rightarrow 0$ limit, we can discard $\bf{I}_4$. The integrals over $\bf{I}_2$ and $\bf{I}_3$ are equal:
\begin{equation}
	\Sigma_2^{(3)}\Big|_{\bf{I}_2}=	\Sigma_2^{(3)}\Big|_{\bf{I}_3} = \left(\Sigma_2^{(1)}\right)^3 \frac{1}{8} \int_a^{\Delta} d^3x\,e^{G_f(x)} \int_\Delta^{\infty} d^3y \,\bigg( e^{G_f(y)-G_f(x-y)}-1\bigg)+\mathcal{O}(a^0)\,.
\end{equation}
Each divergent piece in $\bf{I}_2$, $\bf{I}_3$, and $\bf{I}_4$ admits of an expansion in powers of $\kappa \Delta$, with logarithmic factors also allowed. Since $\Sigma_2^{(3)}$ does not depend on $\Delta$, the $\Delta$ dependencies of $\bf{I}_1$ to $\bf{I}_4$ cancel among each other, meaning that only $(\kappa \Delta)^0$ terms in the divergent pieces contribute to $\Sigma_2^{(3)}$. We can single out these pieces by taking the limit $\kappa \Delta \rightarrow 0$ (while still demanding that $\Delta >> a$). In this limit the integral over $\bf{I}_1$ disappears,\footnote{ For comparison with the RG computation of 2$d$ sine-Gordon theory, we note that this vanishing does not happen in $2d$, where the integral over $\bf{I}_1$ contains a divergent piece proportional to $(\kappa \Delta)^0\log(a/\Delta)$.} and we find that
\begin{equation}
\label{E:mainInew}
	\left(\Sigma_2^{(1)}\right)^{-3}\Sigma_2^{(3)}= \frac{1}{4} \int_a^{\infty} d^3x\,e^{G_f(x)} \int_a^{\infty} d^3y \,\bigg( e^{G_f(y)-G_f(x-y)}-1\bigg)+\mathcal{O}(a^0)\,.
\end{equation}
To evaluate the integral on the right hand side of \eqref{E:mainInew} we expand the integrand over the $y$ coordinate in powers of $x$ and integrate over the angular coordinates. We find
\begin{align}
\begin{split}
\label{E:mainI2}
	\frac{1}{4} \int_{a}^{\infty}&{ d^3x}\, e^{G_f(x)} \int_a^{\infty}  d^3y \left(e^{G_f(y)-G_f(x-y)}-1\right)
	\\ 
	&=
	\frac{2 \pi^2}{3} I_1(a) \int_a^{\infty} dy\, \Bigg(y^2 \left(\frac{\partial G_f}{\partial y}\right)^2 - \frac{\partial}{\partial y} \left( y^2 \frac{\partial G_f}{\partial y} \right)
	\Bigg)
	+ \mathcal{O}(a^0)
	\,.
\end{split}
\end{align}
The total derivative on the right hand side of \eqref{E:mainI2} can be evaluated using \eqref{E:Gfree}
Reinserting the infrared cutoff $\mu$, we find
\begin{equation}
\begin{aligned}
\label{E:totald}
    \int_a^{\infty} dy \frac{\partial}{\partial y} \left( y^2 \frac{\partial G_f}{\partial y} \right) &=  y^2 \frac{\partial G_f}{\partial y} \Big{|}_{y=\infty} -  y^2 \frac{\partial G_f}{\partial y} \Big{|}_{y=a} \\
    &=-\frac{g}{2\pi^2} \frac{\sin(2\pi y \mu)}{g h_r \kappa + 2\pi \mu} \left(1+\mathcal{O}(y^{-1})\right)\Big|_{y =\infty} + \mathcal{O}\left(a\right)
\end{aligned}
\end{equation}
In the $\mu \to 0$ limit this expression is linear in $a$ and, as we will see shortly, will not contribute to the divergences of $\Sigma_2^{(2)}$ for the values of $g$ we are interested in. The same type of behavior is expected if we use a mass regulator for the infrared behavior.\footnote{
With a positive mass regulator $\Sigma_0$, the inverse propagator is $\frac{1}{g}(2\pi k)^3 + h(2\pi k)^2+\Sigma_0$ and the position space propagator $G_{f}(y)$ is obtained by the Fourier transform with no cutoffs. In this case the large $y$ asymptotic of $G_{f}(y)$ is $e^{-y\sqrt{\frac{\Sigma_0}{h}}}/(4\pi h y)$. Thus, the limit $\lim_{y\to\infty}y^2G_f'(y)=0$.
} If we set the infrared cutoff to zero from the start we obtain a constant contribution to the right-hand side of \eqref{E:totald}.

The remaining term in \eqref{E:mainI2} requires more work.
\begin{align}
\begin{split}
	\int_a^{\infty} dy\, y^2 \left(\frac{\partial G_f}{\partial y}\right)^2 &= \frac{g^4 h_r^2 \kappa^2}{4 \pi^4}\int d\rho\, d\sigma\, dx \frac{\rho \sigma \sin(\rho)\sin(\sigma)}{(g h_r \kappa x \rho+\rho^2)^2(g h_r \kappa x \sigma + \sigma^2)^2} \\
	&= \frac{g}{4 \pi^4 h_r \kappa} \int d\rho\, d\sigma \frac{\sin(\rho)\sin(\sigma) (\rho^2 - \sigma^2 + 2 \rho\sigma \ln(\sigma/\rho))}{\rho\sigma(\rho-\sigma)^3}  +\mathcal{O}(a) \\
	& = \frac{g}{4 \pi^4 h_r \kappa} \int d\beta\, d\sigma \frac{\sin(\sigma)\sin(\beta\sigma)(\beta^2-1-2\beta \ln(\beta))}{\beta\sigma^2 (-1+\beta)^3}   +\mathcal{O}(a)  \\
	& = \frac{g}{8 \pi^3 h_r \kappa} + \mathcal{O}(a)\,.
\end{split}
\end{align}
Thus,
\begin{equation}
	\left(\Sigma_2^{(1)}\right)^{-3}\Sigma_2^{(3)} = 	\frac{g}{12 \pi h_r \kappa} \left(I_1(a)-I_1(\Delta)\right) \,.
\end{equation}

It is now straightforward to compute $\Sigma_2^{(3)}$ explicitly. One finds that it contains a logarithmic divergence for $g^2 = 10\pi^2$, power law divergences for $g/2\pi^2 < 5$ and non-integer, and a mix of logarithmic and power law divergences for integer $g/2\pi^2 >5$, reminiscent of \eqref{E:Sigma22e}. Somewhat magically, these divergent contributions are exactly cancelled by the renormalized coupling $h_r$ in \eqref{E:h0order2} when evaluating $\Sigma_2^{(1)}(0)+\Sigma_2^{(3)}(0)$. Thus,
\begin{equation}
\label{E:lambda0}
	\lambda_0 = \lambda_r (\kappa a)^{3-\frac{g}{4\pi^2}} + \mathcal{O}(\lambda_r^5) \hspace{10mm} g\leq 12\pi^2\,.
\end{equation} 
The order $\lambda_r^3$ divergences in $\Sigma_2^{(1)}(0)$ and $\Sigma_2^{(3)}(0)$, which delicately cancel one another, are of order $\frac{1}{h_r}$. Had these terms not cancelled, the $h_r\rightarrow 0$ limit would have been singular, and it would have been unclear whether the purely non-local theory ($h_r=0$) was well-defined. If one undertakes a perturbative analysis of purely non-local sine-Gordon theory, it becomes necessary to introduce an IR regulator $\mu$ as mentioned in the beginning of \ref{sec:SineGordon}. It then turns out that order $\lambda_0^3$ divergences are $\mu$-dependent and can only be eliminated if $\lambda_r$ is a function, not just of $\lambda_0$, $\kappa$, and $a$, but also of $\mu$. In other words, the cancellation of $\lambda_0^3$ divergences cures the purely non-local theory of pathological UV-IR mixing.

\subsection{The $\beta$ functions}
We are now in a position to compute the beta function for the renormalized couplings $\lambda_r$, $h_r$, (and $g_r$). We find:
\begin{align}
\begin{split}
	\beta_g =&\, 0 \, , \\
	\beta_h=& \left.-h_r + \lambda_r^2 \times \begin{cases}
		-\frac{g-10\pi^2}{2\pi^2} C_1(g) -\frac{g-12\pi^2}{2\pi^2} C_2(g) h_r  	&  \substack{\displaystyle g\neq 2 n \pi^2 \\ n=5,\,6,\,\ldots } \\
		-\frac{\pi}{3} + C_3 h_r & g=10\pi^2 \\
		-C_4 -6 \pi^4 h_r & g =12 \pi^2  \\
	\end{cases}\hspace{1mm}\right\}  + \mathcal{O}(\lambda_r^4) \, ,\\
	\beta_{\lambda} =& \left(\frac{g}{4\pi^2}-3 \right) \lambda_r + \mathcal{O}(\lambda_r^5) \qquad 0<g \leq 12 \pi^2 \, .
\end{split}
\end{align}
Some comments are in order. We note that one can choose the $C_i$'s such that the $\beta$ functions are not continuous functions of $g$. This might seem somewhat alarming at first, but we remind the reader that $g$ is an exactly marginal direction in the space of couplings, so there is no flow in the $g$ direction. 

In addition, one may always generate a smooth beta function by an appropriate prescription for finite counterterms. In particular, one can choose $C_1(g)$ and $C_2(g)$ to be functions that are smooth for $g\neq 10\pi^2$ and $g\neq 12\pi^2$ and which satisfy
\begin{align}
\begin{split}
	&C_1(g) \hspace{5mm}\rightarrow  \frac{2\pi^2}{g-10\pi^2} \times \frac{\pi}{3} \text{ as }g\rightarrow 10\pi^2\\
	&C_1(12\pi^2)  = \frac{2\pi^2}{g-10\pi^2}  C_4 \\
	&C_2(10\pi^2)  = -\frac{2\pi^2 }{g-12\pi^2} C_3 \\
	&C_2(g) \hspace{5mm}\rightarrow  \frac{2\pi^2}{g-12\pi^2}\times 6 \pi^4   \text{ as }g\rightarrow 12\pi^2
\end{split}
\end{align}
In particular, one can choose $C_1(g) = \frac{2\pi^2}{g-10\pi^2} \times \frac{\pi}{3}$, $C_2(g) = \frac{2\pi^2}{g-12\pi^2}\times 6 \pi^4$, $C_3 = -6 \pi^4$, $C_4 = \frac{\pi}{3}$ in which case,
\begin{equation}
	\beta_h= -h_r +  \left( 
		-\frac{\pi}{3}  -6 \pi^4 h_r 
		\right)\lambda_r^2
\end{equation}
at least for $g \leq 12\pi^2$. Alternately, one may always choose $C_1$ and $C_2$ so that $\beta_h = -h_r + \mathcal{O}(\lambda_r^4)$ for $g$ sufficiently smaller than $10\pi^2$.

Finally, one may inquire whether it is possible to modify the $-\pi/3$ term in the beta function for $h_r$ at $g=10\pi^2$ or the $-6\pi^4 h_r$ term at $g=12\pi^2$. We claim that no such counterterms are possible. Our argument relies on the Poincare-Dulac theorem (see, e.g., \cite{n61}); both of the aforementioned terms are associated with resonant monomials.

Thus, it seems that $\lambda_r$ is marginal for $g=12\pi^2$ at least up to order $\lambda_r^5$. For $g<12\pi^2$ (and sufficiently small starting value of $\lambda_r$), the sine potential is relevant in the infrared. The coupling to the local kinetic term, $h_r$ cannot be tuned away at $g=10\pi^2$. 
Since $h_r$ is relevant, if it is not tuned to zero, it will dominate the dynamics in the infrared, and the theory will flow towards the three-dimensional theory studied in \cite{Polyakov:1975rs}. We claim that even when $h_r=0$, the infrared theory is confining for $g<12\pi^2$. This is the content of the next section.

\section{Confinement via an area law}
\label{sec:Confinement}

To assess whether the original gauge theory is in a confining phase once a gas of monopoles is included, we may evaluate the Wilson loop. This non-local operator depends on a contour $C$ and will exhibit an area law growth for a large loop in the confining phase. In terms of the holonomy, we write, 
\begin{equation}
	F(C)=\left<\exp(-W(C))\right>= \int DA_\mu \, e^{i \oint_{C}A_{\mu}dx^{\mu}} e^{-S},
\end{equation}
where $C$ is a planar contour orthogonal to the $x^3$ axis.
Recall that in the absence of Wilson loops we have, via \eqref{E:MSum},
\begin{equation}
	\int D A_{\mu}\, e^{-S} \sim \int D\phi\, e^{-S_\phi}\,,
\end{equation}
with
\begin{equation}
	S_{\phi} = \frac{1}{2g} \int d^3x\bigg( \phi D^{\sigma} \phi - 2 g \lambda \cos\phi\bigg)\,,
\end{equation}
where we omitted the higher harmonics.

To obtain a manageable expression for the Wilson loop $W(C)$ we write
\begin{equation}
	\int DA_{\mu}\,e^{-S} \sim \int D\phi \exp\left(-\frac{1}{2g} \int d^3x\, \phi D^\sigma\phi \right) \sum_{N=0}^{\infty} \sum_{\{q_i=\pm1\}} \frac{\lambda^N}{N!} \int \prod_{i=1}^{N} d^3x_i\,e^{i q_i \phi(x_i)}
\end{equation}
and notice that evaluated on monopole configurations with charges $q_i$ we have
\begin{equation}
	e^{i \int_C A_{\mu}dx^{\mu} } =e^{i\int_B F_3 dS^3}= e^{i \sum_{i,q_i}q_i \eta(x_i)}
\end{equation}
where
\begin{equation}
	\eta(x) =\ \int_{B} \frac{(x^\mu-y^\mu)}{2|x-y|^3} \delta^3_\mu d^2y 
\end{equation}
where the integration region is in the interior of $B$ ($\partial B = C$). Thus, after a change of variables $\phi \to \phi+\eta$ we find 
\begin{equation}
\label{E:WCfull}
	F(C) = \int D\phi \exp\left(-\frac{1}{2g} \int d^3x\Big( (\phi-\eta)D^\sigma (\phi-\eta)  - 2 g \lambda \cos \phi\Big) \right)\,.
\end{equation}
The Wilson loop was simple to write in the original variables, but there it is difficult to describe the monopoles. In passing to the dual description, the Wilson loop is the recipe that $\phi$ sees a background connection with a specified monodromy around $C$. In the semi-classical approximation, we can evaluate $W(C)$ by saddle point with these boundary conditions. In practice, for a large area contour we look for field configurations which `jump' between two vacua of $\phi$ in a neighborhood of $B$. This means solving the non-local equation of motion for $\phi$,
\begin{equation}
    D^{\sigma} \phi - D^{\sigma} \eta = -g \lambda \sin(\phi)\,,
\end{equation}
where for $\sigma>2$, we can safely write $D^\sigma$ as $-D^{\sigma-2} \nabla^2$ and then we have 
\begin{equation}
    D^{\sigma} \eta=-D^{\sigma-2}(2\pi \delta'(z)\theta_B(x_{\bot}))
\end{equation}
and
\begin{equation}
\label{E:WCEOM}
	D^{\sigma} \phi + D^{\sigma-2} (2\pi \delta'(z)\theta_B(x_{\bot})) = -g \lambda \sin(\phi)\,,
\end{equation}
with $\theta_B(x)$ equal to unity inside $B$ and zero elsewhere.
The boundary conditions on $\phi$ are that it vanishes at infinity. In particular,
\begin{equation}
	\lim_{z \to \infty} \phi = 0\,.
\end{equation}
In order to solve \eqref{E:WCEOM} it is useful to consider its one-dimensional homogeneous version
\begin{equation}
\label{E:NLSG}
	D^\sigma \phi_0(z) = - g \lambda \sin(\phi_0(z))
\end{equation}
which possesses kink type solutions. While our main interest is $\sigma = 3$, we can also consider other cases. For $\sigma=2$, we have the standard domain wall solution,
\begin{equation}
	\phi_0(z) = 4 \arctan(e^{-\sqrt{g \lambda} z})
\end{equation}
or its anti-kink version $2\pi - \phi_0(z) = \phi_0(-z)$.
As it turns out, the kink solution for $\sigma=1$ can be found explicitly and it takes the simple analytic form
\begin{equation}
	\phi_0(z) = \pi - 2 \arctan(g\lambda z)\,,
\end{equation}
(with $g \lambda$ positive). For other values of $\sigma$ we need to resort to numerics to solve \eqref{E:NLSG}. The dependence of $\phi_0$ on $g\lambda$ can be obtained by a scaling argument. If $\phi_0(z)$ is a solution to \eqref{E:NLSG} with $g\lambda=1$ then $\phi_0(z (g\lambda)^{\frac{1}{\sigma}})$ is a solution to \eqref{E:NLSG} for general $g$ and $\lambda$. Thus, when looking for numerical solutions to \eqref{E:NLSG} we may focus on the $g \lambda=1$ case.

\begin{figure}[hbt]
\begin{center}
\includegraphics[width=0.8\linewidth]{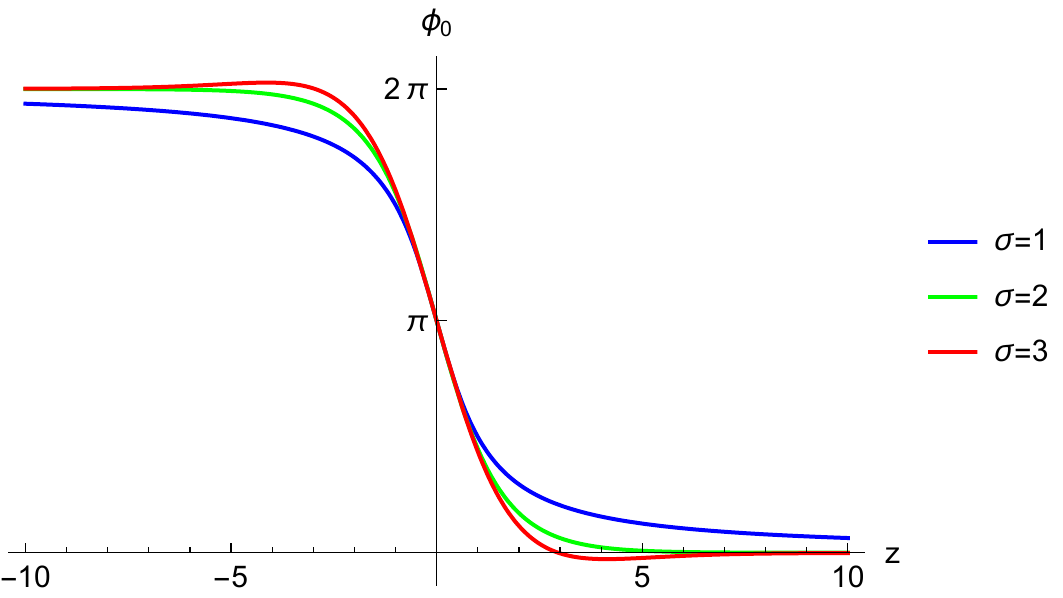}
\end{center}
\caption{\label{F:kink}Solutions to $D^\sigma \phi_0(z) = - \sin(\phi_0(z))$ with $\sigma=1$, $\sigma=2$, and $\sigma=3$. The $\sigma=3$ solution is generated for $N=50$, $\Delta=0.4$, and $S(x)=\frac{1}{2} \left(1-\tanh x  \right)$.}
\end{figure}

Our strategy for solving \eqref{E:NLSG} numerically is to expand $\phi_0$ as a series of sigmoid functions, $S(z)$ (such that $S(z)$ interpolates between $1$ and $0$ and is centered around $z = 0$),
\begin{equation}
	\phi_0(z) = \sum_{n=-N}^N a_n S(z - n \Delta) \,,
\end{equation}
with $\Delta$ a small positive constant, whose value is chosen such as to optimize the fit. One can use, for example,
\begin{equation}
	S(z) = \frac{1}{2} \left(1-\tanh z  \right).
\end{equation}
Other sigmoid functions we have used involve $\arctan(z)$, $1/(1+e^{-z})$ or $\hbox{erf}(z)$. It is straightforward to evaluate the non-local derivative of the sigmoid functions by evaluating them in Fourier space, e.g.,
\begin{equation}
	 D^{\sigma} \left( \frac{1}{2}(1-\hbox{erf}(z) ) \right)= - \frac{2^{\sigma}}{\pi}  \Gamma\left(\frac{1+\sigma}{2}\right) {}_1 F_1\left(\frac{\sigma+1}{2},\,\frac{3}{2},\,-z^2\right) z\,.
\end{equation}
With $D^{\sigma} S(x)$ at hand we can use standard Newton-Raphson methods to search for $a_n$'s such that
\begin{equation}
\label{E:rootsearch}
	\sum_{n=-N}^N a_n D^{\sigma} S(z-n\Delta) + g\lambda \sin\left(\sum_{n=-N}^N a_n S(z-n\Delta)\right) = 0\,,
\end{equation}
is satisfied on $2N+1$ sampling points $z=n\Delta$.
We found that our algorithm becomes more stable if we constrain the $a_n$'s to satisfy
\begin{equation}
	a_n = a_{-n} 
\end{equation}
for $n\neq 0$, and $\phi_0(0) = \pi$ or equivalently
\begin{equation}
	a_0 = 2\Big(\pi-\sum_{n=1}^N a_n\Big)\,.
\end{equation}
A typical numerical kink solution can be found in figure \ref{F:kink}.  

With the homogeneous solution to \eqref{E:NLSG} at hand we can solve \eqref{E:WCEOM} by gluing two such solutions, 
\begin{equation}
\label{E:thesol}
	\phi = \begin{cases}
		\phi_0(z) & z>0,\,\{x,y\} \in B \\
		\phi_0(z)-2\pi & z<0,\,\{x,y\} \in B \\
		0 & \{x,y\} \notin B
	\end{cases}
\end{equation}
valid away from the contour $C$. One can now evaluate \eqref{E:WCfull} in the saddle point approximation,
\begin{equation}
\label{E:Wilsonsaddle}
	F(C) =\left<\exp\left({-W(C)}\right)\right>= \exp\left(-\gamma \int_B d^2y\right)
\end{equation}
where
\begin{equation}
	\gamma = \frac{1}{2 g} \int dz\bigg( \phi_0 D^\sigma \phi_0+2 g \lambda \Big(1-\cos(\phi_0)\Big)\bigg)\,.
\end{equation}
Thus the Wilson loop $W(C)$ in \eqref{E:Wilsonsaddle} satisfies an area law implying that the theory is in a confining phase.

\section{Discussion}
\label{sec:Discussion}

In this paper we considered a generalized free Maxwell theory with Lagrangian $\frac{1}{e^2}F_{\mu\nu}D^{-1}F^{\mu\nu}$ and used electric-magnetic duality to recast the theory in terms of a compact boson with kinetic term $\frac{1}{2g}\varphi D^3 \varphi$. We argued that if the theory is defined on a lattice, then instanton effects generate a cosine potential for the compact scalar. When $g\geq 10\pi^2$ we observed that the non-local sine-Gordon theory is renormalizable only if we introduce an additional (relevant) local kinetic term to the theory. In the generic case when a local kinetic term is present, we expect to be in a confining phase of the theory by the arguments presented in \cite{Polyakov:1975rs,Polyakov:1976fu}.
In the regime where the kinetic term is not generated, or if we fine-tune the theory to set it to zero, we obtain a purely non-local theory. In this theory, we have seen that the cosine potential is an irrelevant deformation for $g>12\pi^2$, while for $g<12\pi^2$ the cosine term is relevant and by the arguments of section \ref{sec:Confinement}, we again expect to be in a confining phase.

\begin{figure}[hbt]
\begin{center}
\includegraphics[width=0.8\linewidth]{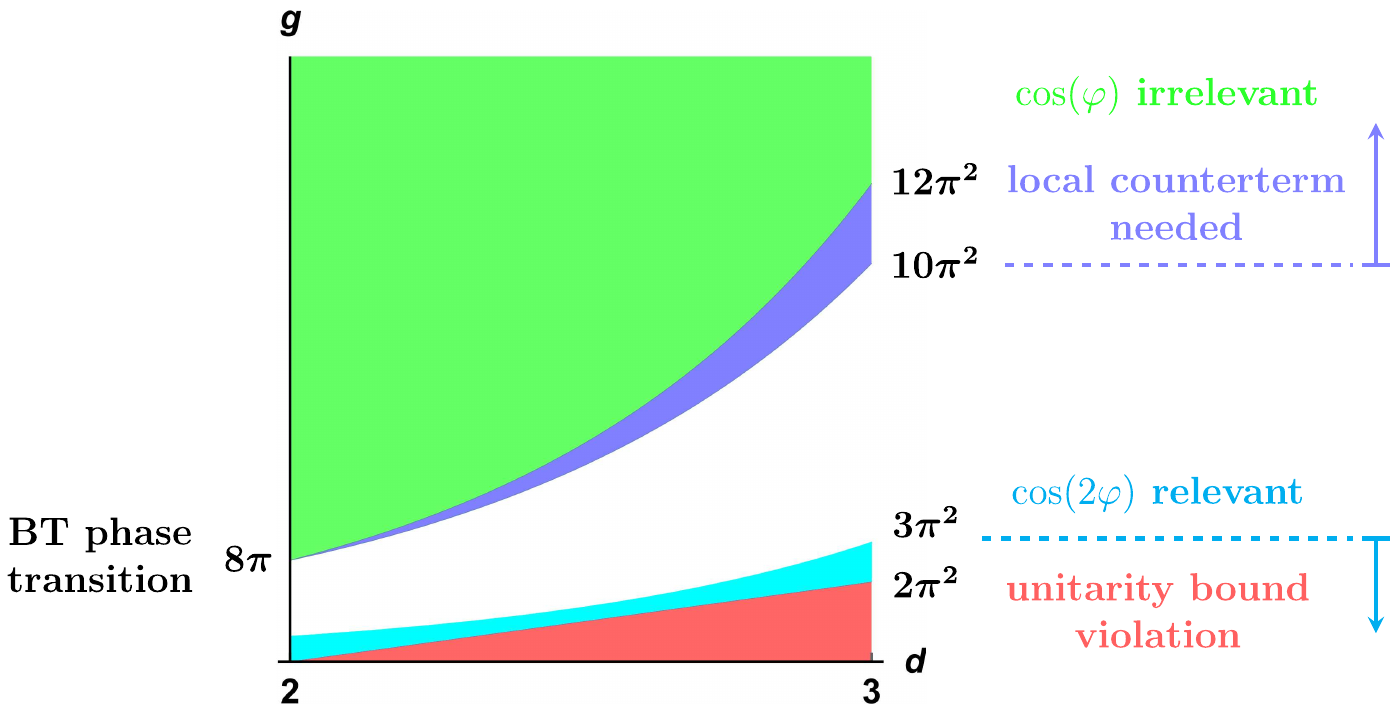}
\end{center}
\caption{Phase diagram of generalized sine-Gordon theory with action \eqref{eq:gendS} for $2<d \leq 3$ with the local kinetic deformation tuned to zero.
}
\label{fig:Overview}
\end{figure}

It is, perhaps, useful to relate the above behavior of the non-local sine-Gordon theory in three dimensions to the behavior of the local sine Gordon theory in two dimensions by analytically continuing in dimensionality, $d$. Indeed, consider the non-local theory of a dimensionless compact scalar, 
\begin{align}
S = \int d^dx \bigg(\frac{1}{2g}\varphi D^d \varphi-\lambda\cos(\varphi)\bigg)\,,
\label{eq:gendS}
\end{align}
where $2 \leq d \leq 3$. In this more general non-local theory, one can show that a local kinetic counterterm is required for $g \geq \frac{d+2}{2}(4\pi)^{d/2}\Gamma(\frac{d}{2})$. When the local kinetic term is tuned away, the cosine is relevant for $g< d(4\pi)^{d/2}\Gamma(\frac{d}{2})$. When $d=2$, these two curves intersect at the value $g=8\pi$, which marks the location of the Berezinskii-Kosterlitz-Thouless phase transition. We plot the different regimes of the purely non-local theory in figure \ref{fig:Overview},  where we also show the region where the potential term $\cos(2\varphi)$ is relevant, $g < \frac{d+2}{8}(4\pi)^{d/2}\Gamma(\frac{d}{2})$, and have excluded the region in which the monopole operator $e^{i\varphi}$ violates the unitarity bound, $g < 2\pi^2(d-2)$. In the figure we have not included the higher harmonics, which are all irrelevant everywhere outside the excluded region in 3$d$.

It may seem puzzling that while in gauge field language the non-local term $F_{\mu\nu}D^{-1}F^{\mu\nu}$ dominates over the standard Maxwell term $F_{\mu\nu}F^{\mu\nu}$ in the IR, in the dual language of compact bosons, the local term $(\partial_{\mu}\varphi)(\partial^{\mu} \varphi)$ dominates over the non-local term $\varphi D^3 \varphi $. One may ask, in the original gauge theory language, what is the relevant operator that deforms $F_{\mu\nu}D^{-1}F^{\mu\nu}$ corresponding to the generation of the local kinetic term in the non-local sine-Gordon theory? To answer this question, we note that by electric-magnetic duality the compact scalar deformation
\begin{align}
\varphi D^3\varphi \rightarrow \varphi \Big(D^3 + h D^2\Big)\varphi 
\end{align}
corresponds to the following deformation of the generalized Maxwell term:
\begin{align}
F_{\mu\nu}D^{-1}F^{\mu\nu} \rightarrow F_{\mu\nu}\frac{1}{D^1+h}F^{\mu\nu}\,.
\label{sub:F}
\end{align}
Taylor expanding the RHS for small $h$, we see that the generalized Maxwell action is initially deformed by the relevant non-local term $h F_{\mu\nu} D^{-2}F^{\mu\nu}$. In fact, retaining  the higher order terms in the Taylor expansion, we see that the monopole background induces deformations by an infinite tower of terms $h^n F_{\mu\nu} D^{-1-n}F^{\mu\nu}$ with $n=1,2,3$... It is a general concern for non-local theories that once locality is abandoned as a principle to constrain the possible terms that can appear in the action, the RG flow may generate uncontrollably many terms. Luckily, the infinite tower of terms resums to the RHS of \eqref{sub:F}, as reflected in the fact that the dual sine-Gordon theory turned out to be renormalizable. 

Finally, we comment on a possible connection to the string picture of confinement. Given the arguments in section \ref{sec:Confinement}, we have established that the purely non-local theory is confining when the cosine term is relevant. The area law of the Wilson loop comes from the semi-classical saddle in which the dual scalar `jumps' when crossing the loop. There has been long standing analytical and numerical evidence that there exists a dual formulation of the confinement mechanism due to effective strings which describe this interface (see \cite{Karliner:1983ab,Caselle:1995fh,Teper:1998te,Caselle:2005xy,Caselle:2016mqu,Athenodorou:2018sab} for a partial list of references). It would be interesting to find an effective string theory which describes this fractional version of Maxwell theory, perhaps along the lines of \cite{Gubser:2019uyf}.

The evidence we have surveyed indicates that generalized free Maxwell theory with its quantum deformations represents a benign theory that may fruitfully be applied as an effective model of the type of four-dimensional systems with photons coupling to a particles on a codimension-one surface that motivated its study. It remains an intriguing possibility to find or engineer natural systems of this kind which exhibit the confinement mechanism and fine-tuned deconfining phase transition that we have analysed.

\section*{Acknowledgements}

The authors are grateful to Shai Chester, Mykola Dedushenko, Igor Klebanov, Petr Kravchuk, Silviu Pufu, Samson Shatashvili, and Edward Witten for helpful comments and illuminating discussions. M.H. is supported by the U.S. Department of Energy Grant DE-SC0009988 and the Institute for Advanced Study. Z.J. is supported by the ERC-COG grant NP-QFT No. 864583
“Non-perturbative dynamics of quantum fields: from new deconfined
phases of matter to quantum black holes”, by the MIUR-SIR grant
RBSI1471GJ, and by the INFN “Iniziativa Specifica ST\&FI”. AY is supported in part by an Israeli Science Foundation excellence center grant 2289/18.

\appendix

\bibliographystyle{JHEP}
\bibliography{monopoles}

\end{document}